\documentclass[]{pasj00}
\newcommand{\beq}{\begin{equation}}
\newcommand{\beqa}{\begin{eqnarray}}
\newcommand{\eeq}{\end{equation}}
\newcommand{\eeqa}{\end{eqnarray}}

\usepackage{times}

\begin{document}

\title{Resistive MHD Simulations of Star-Disk-Jet System}
\author{Miljenko \v{C}emelji\'{c}$^{1,2}$, Hsien Shang$^{1,2}$ and Tzu-Yang Chiang$^{1,2}$ 
}
\affil{%
$^1$Theoretical Institute for Advanced Research in Astrophysics (TIARA), 
National Tsing Hua University, No. 101, Sec. 2, Kuang Fu Rd., 
Hsinchu 30013, Taiwan}
\email{miki@tiara.sinica.edu.tw}
\affil{%
$^2$Institute for Astronomy and Astrophysics, Academia Sinica, P.O. Box
23-141, Taipei 106, Taiwan
}
\email{shang,tychiang@asiaa.sinica.edu.tw}


\maketitle

\begin{abstract}
Stellar magnetosphere and accretion disk interact, and a result should be 
outflow launched from the innermost vicinity of a 
protostellar object. We simulated physical conditions in this region by resistive MHD
simulations. Outflows resembling the observed ones do not happen in the closest 
vicinity, except for quasi-stationary funnel flows onto the star, but could occur 
at few tens of stellar radii above the star. Numerical simulations we performed 
using our resistive version of ZEUS-3D code, ZEUS347.
\end{abstract}

\section{Introduction}
Astrophysical jets are often present phenomena
in both stellar and galactic scale. In our numerical simulations, we
investigate the interaction of protostellar magnetic field with the large scale 
magnetic field of the circumstellar disk.
\begin{figure}
\FigureFile(80mm,40mm){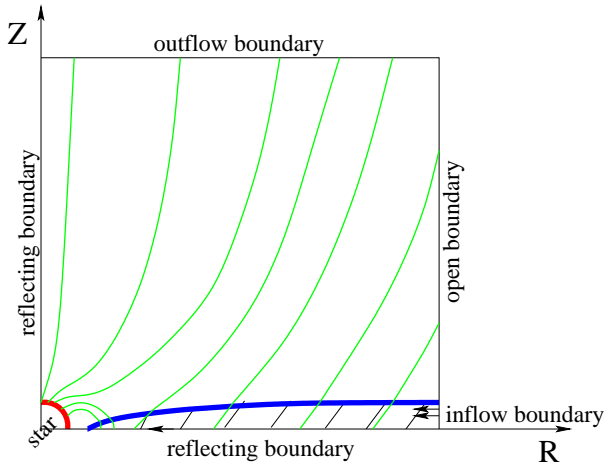}
\caption{Schematic view of our model. A stellar dipole magnetic field,
combined with the disk large scale open magnetic field, threads the disk
and its corona. Boundary conditions for the star-disk simulation are also shown.}
\label{boxtia}
\end{figure}
Numerical simulations of the ideal MHD jet propagation with the disk as a
boundary condition have been presented in Ustyugova et al. 1995. 
They were also studied,
in various setups, in the papers by Ouyed \& Pudritz (1997a,b) and
in the resistive MHD setup in Fendt \& \v{C}emelji\'{c} (2002).

Simulations involving the underlying disk have been firstly presented
for short lasting simulations in Shibata \& Uchida (1985). 
\begin{figure}
\FigureFile(60mm,60mm){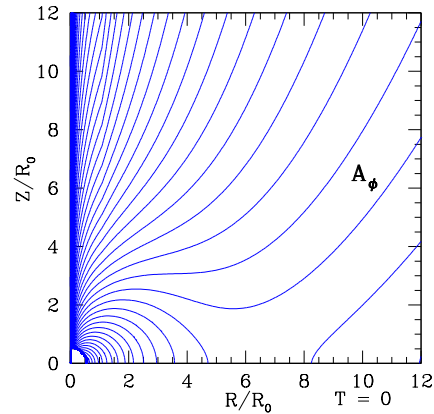}
\caption{Lines of poloidal magnetic field. In the most general configuration
 they are set as a combination of stellar dipole magnetic field and the open 
large scale disk field.
}
\label{a3t0}
\end{figure}    
After success of Casse \& Keppens (2002) to
simulate the jet launched from resistive disk around the protostar during 
more periods ({\em few tens}) of rotations, more investigations on the
effects of resistivity have been done (Zanni et al., 2004; 2007). 
Our setup extends the setup of Casse \& Keppens (2002) (see
also \v{C}emelji\'{c}\ \& Fendt, 2004) for a case when also the stellar
magnetosphere is included in the simulations. A star has been included as a rotating
 sink boundary for the matter, emulating the stellar accretion.

\section{Problem setup}
We solved the resistive MHD equations using the ZEUS347 code in the 
axisymmetry option in cylindrical coordinates. In the time evolution 
of energy equation we neglected the Ohmic part.

The energy of initial state was computed by the polytropic equation of 
state $p$=$K\rho^\gamma$ with $\gamma$=5/3, when
 the internal energy (per unit volume) is defined as $e=p/(\gamma-1)$.
Our simulations presented here have been done in a resolution R$\times$Z=
(320$\times$320)
grid cells. The physical scale has been typically
R$\times$Z=(10$\times$10) stellar radii. Simulations in smaller and larger
resolutions and scales have also been performed.

The initial disk corona in a hydrostatic equilibrium, co-rotating with the
underlying disk, has been set. The central star was
considered as a (rotating or non-rotating) sink for matter inflowing from 
the disk. The disk
itself has been given as rotating with the slightly sub-Keplerian rotation profile.
Sketch of our model is shown in Fig. \ref{boxtia}.
In the most general setup the magnetic field has been set as a stellar dipole 
combined with split-monopole large scale field,
threading the disk, as shown in Fig. \ref{a3t0}. For the investigation of the 
innermost part of the magnetosphere, we also considered only the stellar 
dipole field case.

The magnetic diffusivity is essential for lifting of the matter from the disk,
and it has been introduced with a Gaussian profile, depending on height above
the disk equator. It has been parametrized by the local Alfven velocity, and it
 was effectively zero outside the disk.
\section{Results}
\begin{figure}
\FigureFile(70mm,40mm){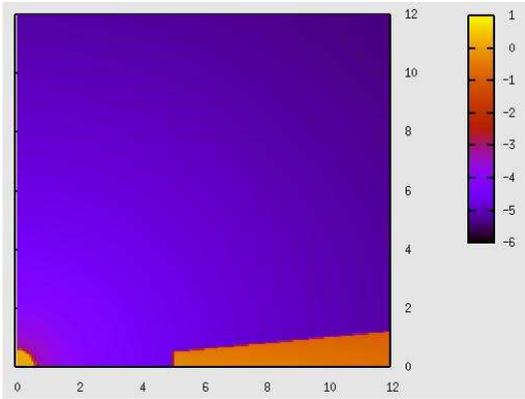}
\caption{Computational box in our simulations. Initial setup for a 
disk-star simulation. Density shown in color
grading.
}
\label{initdens}
\end{figure}

\begin{figure}
\FigureFile(70mm,40mm){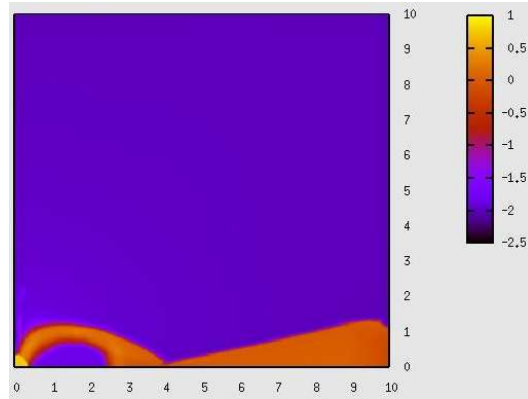}
\caption{For the close vicinity of a star, solutions with the funnel of matter
from the disk onto the star are expected. Conditions for such solutions and
their stability are still under investigation.
}
\label{infall}
\end{figure}

In our simulations here we present the results for innermost part of the star-disk
system. A magnetic field configuration determines the time evolution of initial 
configuration. In Fig. \ref{initdens} shown is the initial density distribution in 
our computational box. Following in Fig. \ref{infall} is the solution 
after few rotations at the inner disk radius.
\begin{figure}
\FigureFile(60mm,35mm){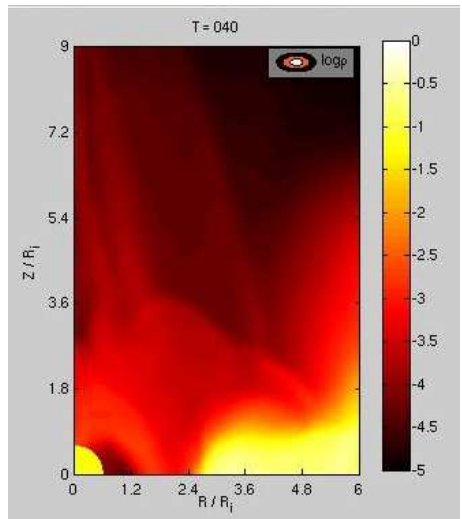}
\caption{After a few ten rotations at the inner disk radius, the new equilibrium
is established. Such configuration could, at larger distances from the protostar,
feed a collimated outflow. In this simulation, shown after T=40 rotations, the 
inner disk radius initially was R$_{\rm in}$=3.
}
\label{denst20}
\end{figure}
Our results are comparable with these of Romanova et al. (2002) and 
Long et al. (2005), when an infall onto the star is observed.

The disk setup is still much simplified, representing rather a clump of matter 
in a hydrostatic balance than an accretion disk. However, we can study
 the interaction of such disk and star, through
the difference in the rotation rates and magnetic field configurations of stellar
and disk fields. A perspective of fully 3D simulations for such systems closes 
us to more realistic theoretical investigations of a star-disk interaction.

\end{document}